\documentclass[12pt]{article}
 \usepackage{epsfig}
 \def\be{\begin{equation}}
 \def\ee{\end{equation}}
 \def\bea{\begin{eqnarray}}
 \def\eea{\end{eqnarray}}
 \usepackage{graphicx}

 \catcode`\@=11
 \def\lsim{\mathrel{\mathpalette\@versim<}}
 \def\gsim{\mathrel{\mathpalette\@versim>}}
 \def\@versim#1#2{\vcenter{\offinterlineskip
 \ialign{$\m@th#1\hfil##\hfil$\crcr#2\crcr\sim\crcr } }}
 \catcode`\@=12

 \catcode`@=12
\begin{document}
 \thispagestyle{empty}
 \begin{flushright}
 UCRHEP-T629\\
 Feb 2025\
 \end{flushright}
 \vspace{0.6in}
 \begin{center}
 {\LARGE \bf Scotogenic Froggatt-Nielsen and the\\ 
Versatility of Soft Symmetry Breaking\\}
 \vspace{1.5in}
 {\bf Ernest Ma\\}
 \vspace{0.1in}
{\sl Department of Physics and Astronomy,\\ 
University of California, Riverside, California 92521, USA\\}
 \vspace{1.2in}

\end{center}

\begin{abstract}\
Preserving the unique role of the one Higgs doublet of the standard model, 
it is proposed that quark and lepton mass patterns, often ascribed to the 
Froggatt-Nielsen mechanism using nonrenormalizable higher-dimensional 
terms, may be enforced in a \underline{renormalizable} theory of just one 
Higgs doublet by the scotogenic mechanism with soft symmetry breaking in 
the dark sector. A revised version of the original $A_4$ model of charged 
leptons and neutrinos is discussed.
\end{abstract}

\newpage

\baselineskip 24pt
\noindent \underline{\it Introduction}~:~ 
The standard model (SM) of particle interactions requires only one Higgs 
doublet for all quark and lepton masses. (This applies also to neutrinos 
if they are Dirac fermions, whereas Majorana neutrino masses could have a 
variety of different origins~\cite{m98}.) However, it is often assumed that 
there may be other scalars, doublets and singlets, in order to support an 
underlying symmetry for understanding the three families of quarks and 
leptons, with their accompanying mass and mixing patterns.

Broadly speaking, there are two well studied approaches. One is to have two 
or more Higgs doublets coupling selectively to the quark and leptons in a 
renormalizable theory. If a family symmetry is also imposed, the Higgs sector 
may be rather complicated and flavor changing interactions are usually 
inevitable.  Another is to retain only one Higgs doublet, but to allow 
nonrenormalizable higher-dimensional Yukawa operators involving neutral 
scalar singlets (flavons) carrying appropriate new charges matching those of 
the quarks and leptons, known widely as the Froggatt-Nielsen (FN) 
mechanism~\cite{fn79}.  In this paper, it will be shown how the latter may be 
implemented simply in a \underline{renormalizable} theory, utilizing the 
scotogenic framework~\cite{m06,t96} with soft symmetry breaking.

Recognizing the versatility of this new idea, the original $A_4$ model of 
charged leptons is recast where the three Higgs doublets are replaced by 
just the one SM Higgs doublet with the help of dark singlet flavons.  The 
corresponding neutrino sector will also be discussed.

\noindent \underline{\it Example}~:~
In the SM, under the gauge symmetry $SU(2)_L \times U(1)_Y$, left-handed 
fermions are doublets and right-handed fermions are singlets. They are 
connected by the one Higgs doublet $\Phi = (\phi^+,\phi^0)$ and obtain 
Dirac masses from the vacuum expectation value $\langle \phi^0 \rangle = v$.
Consider for example $(u,d)_L, (c,s)_L$, and $u_R,c_R$. In the SM, the 
$2 \times 2$ Yukawa coupling matrix linking $(\bar{u},\bar{c})_L$ to 
$(u,c)_R$ is proportional to $\bar{\phi}^0 = (v + h/\sqrt{2})$, where 
$h$ is the one residual physical Higgs boson after the spontaneous symmetry 
breaking of $SU(2)_L \times U(1)_Y$ to $U(1)_Q$.  Hence the diagonalization of 
the $2 \times 2$ mass matrix also leads to the diagonalization of the $h$ 
couplings, i.e. $(h/v\sqrt{2})(m_u \bar{u}u + m_c \bar{c}c)$.

To understand why $m_u << m_c$, a pssible approach is to postulate a new 
global $U(1)$ symmetry, under which
\begin{equation}
(c,s)_L,c_R \sim 0, ~~~ (u,d)_L \sim 1, ~~~ u_R \sim -1,
\end{equation}
together with a scalar singlet flavon $\eta \sim 1$, so that the allowed 
Yukawa couplings are
\begin{equation}
\bar{c}_Lc_R \bar{\phi}^0, ~~~ \bar{c}_L u_R \bar{\phi}^0 (\eta/\Lambda), ~~~ 
\bar{u}_L c_R \bar{\phi}^0 (\eta/\Lambda), ~~~ \bar{u}_L u_R \bar{\phi}^0 
(\eta/\Lambda)^2,
\end{equation}
where $\Lambda$ is an arbitrary mass-dimensional parameter. This is the FN 
mechanism in a nutshell.  Allowing $\eta$ to have a vacuum expectation 
value introduces a small parameter $\langle \eta \rangle/\Lambda$ which 
explains $m_u << m_c$.

\noindent \underline{\it Tree-Level Ultraviolet Completion}~:~ 
To convert the above example to that of a renormalizable theory, consider the 
addition of two heavy quark singlets $x,y$ with the same electric charge as 
$u,c$.  Let their FN charges be
\begin{equation}
x_{L,R} \sim 1, ~~~ y_{L,R} \sim 0,
\end{equation}
then the $4 \times 4$ mass matrix linking $(\bar{u},\bar{c},\bar{x},\bar{y})_L$ 
to $(u,c,x,y)_R$ is given by
\begin{equation}
{\cal M} = \pmatrix{0 & 0 & m_{ux} & 0 \cr 0 & m_c & 0 & m_{cy} \cr 
0 & \epsilon_{xc} & M_x & \epsilon_{xy} \cr \epsilon_{yu} & 0 & 
 \epsilon_{yx} & M_y},
\end{equation}
where $m_{ux},m_c,m_{cy}$ come from $\langle \phi^0 \rangle$, $\epsilon_{xc},
\epsilon_{xy}, \epsilon_{yu}, \epsilon_{yx}$ come from $\langle \eta \rangle$, 
and $M_x,M_y$ are allowed invariant masses. The induced $\bar{u}_L c_R, 
\bar{c}_L u_R$, and $\bar{u}_L u_R$ masses are then
\begin{equation}
m_{uc} = {m_{ux} \epsilon_{xc} \over M_x}, ~~~ 
m_{cu} = {m_{cy} \epsilon_{yu} \over M_y}, ~~~ 
m_{uu} = {m_{ux} \epsilon_{xy} \epsilon_{yu} \over M_x M_y} - 
{m_{uc} m_{cu} \over m_c},
\end{equation}
exactly as expected in the FN mechanism.
This analysis also shows that there must be mixing between the $(u,c)$ 
and $(x,y)$ sectors.  Hence flavor changing neutral interactions are 
unavoidable.

\noindent \underline{\it Scotogenic Realization}~:~
Let there be a dark $Z_2$ symmetry where all SM particles are even, including 
the one Higgs doublet, with the FN charges of Eq.~(1) for the purpose of 
the example being discussed. The dark sector where all the particles are odd 
includes one quark doublet $(a,v)_{L,R}$ and two 
singlets $a'_{L,R},v'_{L,R}$ with charges $(2/3,-1/3)$, all having no FN charge. 
There are also two dark scalar singlets, $\eta \sim 1$ as before and 
$\zeta \sim 0$.  The scotogenic realization of Eq.~(2) is then achieved 
via Fig.~1.
\begin{figure}[htb]
\vspace* {-3.5cm}
\hspace*{-3cm}
\includegraphics[scale=1.0]{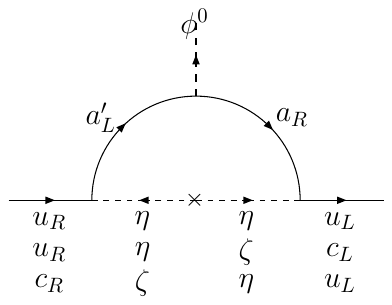}
\vspace{-22.0cm}
\caption{Scotogenic Froggatt-Nielsen.}
\end{figure}
 
It should be emphasized that $\eta$ and $\zeta$ belong to the dark sector and 
have no vacuum expectation values, in contrast to the previous scenarios. 
The soft breaking of the FN symmetry occurs between the internal scalar lines 
and is confined within the dark sector. There are no tree-level flavor 
changing neutral interactions in the SM.  However, since fermion masses 
are of radiative origin (at least partially), anomalous Higgs couplings 
will appear as first shown in Ref.~\cite{fm14}.  The generalization of this 
approach for any conventional FN model is obvious.  The dark quarks 
should be heavy, say of order few TeV.

\noindent \underline{\it Revised $A_4$ Model of Charged Leptons}~:~
In the well-known original $A_4$ model~\cite{mr01} of charged leptons and 
neutrinos, the three left-handed lepton doublets, the three right-handed 
charged lepton singlets transform as
\begin{equation}
L_{iL} = (\nu_i,l_i)_L \sim \underline{3}, ~~~ l_{iR} \sim \underline{1}, 
\underline{1}', \underline{1}''.
\end{equation}
They are connected by three Higgs doublets
\begin{equation}
\Phi_i = (\phi_i^+,\phi_i^0) \sim \underline{3}.
\end{equation}
As $A_4$ breaks to $Z_3$ through $\langle \phi_i^0 \rangle = v/\sqrt{3}$, the 
three charged leptons obtain independent masses $m_e,m_\mu,m_\tau$, and the 
unitary transformation linking them to the neutrino mass matrix is
\begin{equation}
{U}_\omega = {1 \over \sqrt{3}} \pmatrix{1 & 1 & 1 \cr 1 & \omega & 
\omega^2 \cr 1 & \omega^2 & \omega},
\end{equation}
where $\omega = \exp{2\pi i/3} = -1/2 + i\sqrt{3}/2$. 
As such, the residual triality symmetry~\cite{m10} distinguishes the three 
lepton families and forbids processes such as $\mu \to e \gamma$.

Utilizing the scotogenic mechanism, the three Higgs doublets may be replaced 
by just the one SM Higgs doublet and a dark sector consisting of
\begin{equation}
(N,E)_{L,R} \sim \underline{1}, ~~~ E'_{L,R} \sim \underline{1}, ~~~ 
\eta_{1,2,3} \sim \underline{1},\underline{1}',\underline{1}'', ~~~ 
\zeta_{1,2,3} \sim \underline{3}.
\end{equation}
The resulting loop diagram is shown in Fig.~2.
\begin{figure}[htb]
\vspace* {-3.5cm}
\hspace*{-3cm}
\includegraphics[scale=1.0]{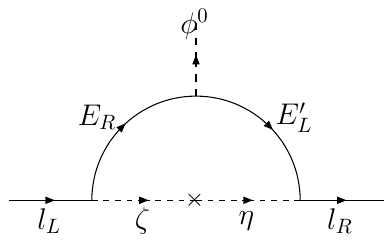}
\vspace{-22.5cm}
\caption{Scotogenic $A_4$ realization.}
\end{figure}

The soft breaking is assumed to be of the form $U_\omega$ times a diagonal 
matrix, so that the 
residual $Z_3$ symmetry is retained which also serves to protect its 
stability.  This scenario mimics that of Ref.~\cite{m15} where the dark 
scalars are charged singlets instead. 

\noindent \underline{\it Neutrino Sector}~:~
The scotogenic neutrino sector is a variation of that in Ref.~\cite{m16}. 
The new feature here is the inclusion of a light dark fermion singlet $S_L$.  
The resulting one-loop diagram for the Majorana 
neutrino mass matrix is shown in Fig.~3.
\begin{figure}[htb]
\vspace* {-5.5cm}
\hspace*{-3cm}
\includegraphics[scale=1.0]{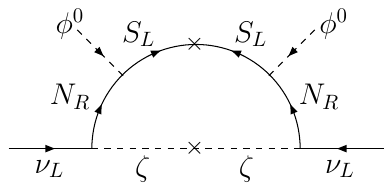}
\vspace{-21.5cm}
\caption{Scotogenic Neutrino Sector.}
\end{figure}

If $A_4$ is exact, then $\zeta_{1,2,3}$ all have the same mass.  The soft 
breaking of $A_4$ allows different masses and each $\zeta_i$ to be a linear 
combination of these mass eigenstates.  As shown in Ref.~\cite{m15,m16}, 
cobimaximal neutrino mixing ($\theta_{23}=\pi/4, \delta_{CP}=\pm \pi/2$) 
is obtained if $\zeta_i$ are real (i.e. not complex) scalars.

\noindent \underline{\it Light Dark Matter}~:~ 
The $3 \times 3$ mass matrix spanning $(\bar{N}_R,N_L,S_L)$ is of the form
\begin{equation}
{\cal M}_N = \pmatrix{0 & m_N & m_1 \cr m_N & 0 & m_2 \cr m_1 & m_2 & m_S},
\end{equation}
where $m_N$ is an invariant mass, $m_1$ comes from 
$\langle \bar{\phi}^0 \rangle$, 
$m_2$ comes from $\langle \phi^0 \rangle$, and $m_S$ is assumed small.   
Hence the coupling of SM Higgs boson to $SS$ is
\begin{equation}
f_h = {m_N \over v \sqrt{2}} \left( {m_1 \over m_N} \right)
\left( {m_2 \over m_N} \right),
\end{equation}
and its decay rate is
\begin{equation}
\Gamma_h = {f_h^2 m_h \over 8 \pi} \sqrt{1-4r^2} (1-2r^2),
\end{equation}
where $r = m_S/m_h$. Now $S$ may be produced by $h$ in a freeze-in scenario 
with the correct relic abundance if~\cite{ac13}
\begin{equation}
f_h \sim 10^{-12} r^{-1/2}.
\end{equation}
This implies $m_S m_1^2 m_2^2/m_N^2 \sim 7.6 \times 10^{-18} ({\rm GeV})^3$.  Let 
$m_N = 2$ TeV, $m_1=m_2= 30$ MeV, then $m_S \sim 38$ keV.

\noindent \underline{\it Concluding Remarks}~:~ 
The idea that family structure of quarks and leptons may be due to the dark 
sector and supported by just the one Higgs doublet of the standard model 
was originally proposed in Ref.~\cite{m14}.  In this paper, the 
Froggatt-Nielsen mechanism is included, emphasizing the role of singlet 
scalar flavons which are now relegated to the dark sector.  The scotogenic 
realization of flavor symmetry is advocated, supported by the soft 
breaking mass-squared terms of dark scalars.  The original $A_4$ model 
of charged leptons and neutrinos is revised in this context.

\noindent \underline{\it Acknowledgement}~:~
This work was supported in part by the U.~S.~Department of Energy Grant 
No. DE-SC0008541.  

\baselineskip 18pt
\bibliographystyle{unsrt}

\end{document}